# Watching Stars in Pixels: The Interplay of Traffic Shaping and YouTube Streaming QoE over GEO Satellite Networks

Jiamo Liu*, David Lerner§, Jae Chung§, Udit Paul*,
Arpit Gupta*, Elizabeth Belding*

University of California Santa Barbara*  ViaSat§

## ABSTRACT

Geosynchronous satellite (GEO) networks are a crucial option for users beyond terrestrial connectivity. However, unlike terrestrial networks, GEO networks exhibit high latency and deploy TCP proxies and traffic shapers. The deployment of proxies effectively mitigates the impact of high network latency in GEO networks, while traffic shapers help realize customer-controlled data-saver options that optimize data usage. It is unclear how the interplay between GEO networks' high latency, TCP proxies, and traffic-shaping policies affects the quality of experience (QoE) for commonly used video applications. To fill this gap, we analyze the quality of over 2 k YouTube video sessions streamed across a production GEO network with a 900 Kbps shaping rate. Given the average bit rates for the selected videos, we expected seamless streaming at 360p or lower resolutions. However, our analysis reveals that this is not the case: 28% of TCP sessions and 18% of gQUIC sessions experience rebuffering events, while the median average resolution is only 380p for TCP and 299p for gQUIC. Our analysis identifies two key factors contributing to sub-optimal performance: (i) unlike TCP, gQUIC only utilizes 63% of network capacity; and (ii) YouTube's imperfect chunk request pipelining. As a result of our study, the partner GEO ISP discontinued support for the low-bandwidth data-saving option in U.S. business and residential markets to avoid potential degradation of video quality—highlighting the practical significance of our findings.

## 1 INTRODUCTION

Geosynchronous satellite (GEO) networks, through providers such as ViaSat and HughesNet, are a key last-mile Internet access technology in challenging environments such as rural and other hard-to-reach communities, aircraft, sea ships, and others. While recently low Earth orbit (LEO) satellite networks have grown in availability, the high cost of deploying and maintaining LEO networks can hinder access for under-served communities [25], making GEO networks an attractive alternative. GEO satellites orbit 22, 000 miles above the earth and move at the same speed as the earth, ensuring that their location remains fixed relative to the ground stations over time. While convenient for routing simplification, the geosynchronicity comes at the cost of high latency; round trip times through GEO satellites are typically 500-600 ms. To mitigate the effects of this latency, GEO ISPs often employ a variety of techniques. These usually include TCP Performance Enhanced Proxies (PEP) with geo-optimized configurations. Additionally, many wireless plans, both terrestrial and satellite, provide customers with a fixed high-speed data quota (gigabytes) per month. After a customer consumes this data, their traffic is deprioritized, which can result in slow speeds for the user when the network is congested. To avoid data deprioritization, customers are typically provided with an option to reduce general data consumption by shaping their video traffic, thereby enabling high-speed data to last longer.

As the dominant Internet application, accounting for approximately 53% [5] of total Internet traffic, video streaming is a critical application to support all network types. Particularly in remote areas, video streaming can be critical for activities such as online education and work. The numerous studies of video streaming QoE are typically conducted either in emulated or production terrestrial networks characterized by low delay and high bandwidth (e.g. [11, 29]). Amongst the results, these studies show that QoE degradation events are rare in terrestrial networks. Prior studies have also analyzed the impact of the transport layer in video performance, in particular comparing TCP and QUIC, the latter of which has gained wide adoption by applications such as YouTube [3, 6, 26, 28]. Some studies have shown that TCP and QUIC have similar performance for video streaming tasks in terrestrial networks [3, 18, 28], while in [13] Google claims that QUIC improves YouTube QoE key performance indicators (KPIs). However, because of their inherent differences, it is not clear whether these results hold in GEO networks, particularly when traffic shaping is employed.

We address this understanding gap through an extensive study of video streaming performance over an operational GEO satellite network. We focus our study on YouTube because of its widespread popularity; current data places YouTube video consumption as outpacing that of Netflix worldwide.[1] We stream 2,080 180-190 second videos, using either the TCP or Google QUIC (gQUIC) protocol, and analyze the resulting QoE KPIs to characterize the video performance. Critically and surprisingly, our first key discovery is that neither TCP nor gQUIC achieves seamless streaming at an average resolution of 360p when traffic is shaped at 900 Kbps, the rate offered to customers in our production network. Specifically, we observe that 28% of TCP sessions and 18% of gQUIC sessions experience rebuffering events, while the median average resolution is 380p for TCP and only 299p for gQUIC.

To understand this observation, we deeply analyze the video traffic and determine that TCP is capable of utilizing all available link capacity during transmission, while gQUIC is only able to utilize 63% of the capacity. Our results suggest that the performance of gQUIC, specifically in conjunction with the BBR congestion control algorithm, is suboptimal when used in GEO networks. Furthermore, both TCP and gQUIC suffer from imperfect chunk request scheduling, resulting in idle time that reduces the overall throughput for TCP by 40% and gQUIC by 30%.

As a result of our study, our partner GEO ISP discontinued support for the low-bandwidth data-saving shaping option in its U.S. business and residential network to avoid potential degradation of video quality. Further, we encourage YouTube and other content providers to more fully consider the operational environment of GEO networks and optimize players to deliver high-quality video despite the presence of high latency links and other GEO network features. Additionally, optimization is needed for the gQUIC + BBR [13] stack to achieve performance comparable to PEP-enabled TCP + BBR in the GEO satellite network.

## 2 BACKGROUND AND MOTIVATION

In this section, we provide background on the three key concepts in this paper. First, we describe video streaming, and in particular the use of adaptive bit rate algorithms and the KPIs that are used to measure video QoE. The characteristics of GEO satellite networks are then discussed, along with optimizations incorporated to provide customers with improved performance. Additionally, we discuss key features of the QUIC protocol, including a discussion of why QUIC is not necessarily the better protocol for the GEO satellite network despite its apparent suitability for video streaming.

### 2.1 Video streaming applications (VSAs)

Internet video streaming services typically divide a video into smaller segments called chunks. These chunks are often of different playback durations; hence the chunks can be of variable size [20]. The video quality is determined by the number of pixels in each frame (i.e., resolution) and the (average) number of bits per second of playback (i.e., bit rate). Most streaming service providers use variable bit rates (VBR) to encode the video into a sequence of frames. The number of bits needed to encode a specific chunk depends on the video type and its quality [16]. In general, high-action/high-resolution chunks require more bits to encode than motionless and/or lower-resolution chunks.

**Adaptive bit rate algorithms.** To optimize the viewing experience, each client maintains a buffer where it stores received chunks. With a repository of chunks, the likelihood of continuous playback during a video session is greatly increased. At the start of the video session, the client waits to fill this buffer to a predefined level before the video playback begins. The client begins by sending an HTTPS request to retrieve a specific segment at a pre-selected quality (240p, 360p, etc.). On receiving this request, the server sends the requested segment to the client.

Each client uses an application-layer *adaptive bit rate (ABR)* algorithm to determine the quality of the request in the next segment. The ABR algorithms employed by most video streaming services are proprietary, but previous work has shown that these algorithms typically use parameters such as estimated bandwidth and current buffer size to determine the quality of the next requested segment [17, 34].

**Quality of experience for VSAs.** Video stream QoE is determined by several KPIs. These include initial buffering time, resolution, and the number of rebuffering events. During a rebuffering event, video playback is paused while the received video is placed into the playback buffer. For optimal QoE, rebuffering events, resolution switches, and initial buffering time should be minimized [27]. A higher resolution is preferred as long as there is sufficient network bandwidth to deliver the video chunks before the playout deadline [10].

The goal of the ABR algorithm is to select the appropriate resolution for each chunk to maximize viewing resolution while also minimizing events that lower QoE. Most video streaming applications work very well in high-bandwidth (more than 10-15 Mbps), low-latency (few tens of ms) networks [24]. However, it is far less clear how well they perform in high-latency networks and shaped bandwidth, which are typical of GEO networks. Further, it is not well-understood how well these applications are able to interact with additional network components, such as TCP proxies and traffic-shaping algorithms, that are common in GEO networks. Hence, it is in these environments that our work focuses.

### 2.2 Geosynchronous satellite networks

Geosynchronous satellite networks use wireless links between ground stations and space satellites to connect subscribers to the Internet. Transparent TCP proxies, which speed up TCP slow-start and congestion recovery, are often employed to mitigate the effects of the long round-trip propagation delays of GEO satellite links. Two TCP proxies are typically used: upstream and downstream. The upstream proxy runs at the ground station, while the downstream proxy runs at the satellite modems closer to the end-users. As a result of these proxies, each video streaming session entails three independent TCP connections: server-to-upstream-proxy (C1), upstream-to-downstream-proxy (C2), and downstream-proxy-to-client (C3), as shown in Figure 1. The upstream proxy acknowledges packets coming from the video server aggressively and therefore increases the congestion window quickly. On the client side, packets are acknowledged immediately as well.

---
[1]For instance, one study states that, in 2022, YouTube represented 15% of traffic on consumer broadband networks, while Netflix represented 9% [5].

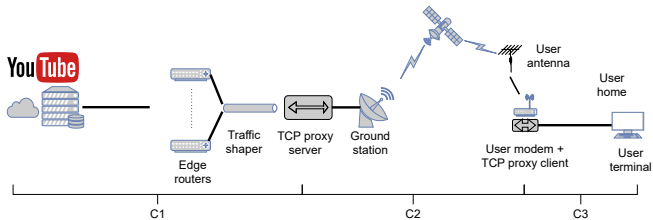

**Figure 1: Testbed configuration.**

In addition to TCP proxies, many satellite and mobile wireless network operators provide data-saver options that use traffic shapers to constrain the bandwidth allocated to different applications [14]. This enables users to view more hours of video or engage in other network activities while reducing the likelihood they exceed their monthly data limit.

**QUIC in GEO satellite networks:** QUIC is rapidly becoming the default transport layer protocol for many video streaming services. YouTube, for example, is predominantly transmitted over QUIC. QUIC runs in user space, uses UDP for data transmission, and is designed for applications that use multiple HTTPS multiplexed connections. Compared to TCP, it expedites the connection setup by reducing the number of round trips required to establish a connection between the two end hosts. During packet losses, QUIC avoids head-of-line blocking by servicing unaffected streams while recovering lost packets. TCP proxies cannot be leveraged by QUIC-based connections because QUIC connections are cryptographically signed end-to-end. Hence the feedback loop of QUIC is delayed in GEO networks. As a result of these nuances, the use of QUIC instead of TCP over GEO links should not be taken for granted, as it is unclear how video QoE will be impacted when used with other components such as PEP and traffic shaper. This observation serves as the driving force behind our investigation.

## 3 METHODOLOGY AND DATASET

In this section, we present the configuration of our testbed and the methodology used to stream YouTube videos while collecting HTTP logs and QoE KPIs. Additionally, we provide a summary of the dataset and experiment metrics collected.

**Testbed.** Figure 1 illustrates the primary components of our testbed network architecture, including the YouTube server, traffic shaper, server-side TCP proxy, satellite link, client-side proxy, and client laptop. Note that because this is a production network, the client laptop is the only component over which we have direct control. To improve TCP performance over the long latency satellite link, TCP traffic is split into three separate connections (C1, C2, and C3), as shown in Figure 1. On the other hand, QUIC traffic is not split into separate connections. The client laptop is used to collect HTTP logs. The satellite provider uses a token bucket traffic shaper to limit the throughput of video traffic on the low-latency link between the YouTube server and the upstream proxy, with a bandwidth shaping rate of 0.9 Mbps and bursts up to 0.99 Mbps. [14] shows that multiple GEO satellite network ISPs utilize this shaping rate. Note that traffic shaping is a subscriber opt-in feature for the ISP in the study.

**YouTube experiments.** We analyzed an open video catalog of approximately 8 million entries released by YouTube [1] to collect representative data from a variety of video types. We randomly selected 13 videos from each of the 16 distinct video categories to form a total pool of 208 videos. We streamed each video ten times: five with QUIC enabled and five with QUIC disabled, for a total of 2,080 sessions; TCP was utilized in the QUIC-disabled sessions. Each video is between 180-190 seconds in length to improve the scalability of the experiments while simultaneously being long enough to allow the congestion window to saturate at its maximum capacity [12]. The average bit rate of each of the 208 videos at 360p and 480p is represented as a CDF in Figure 2. The figure demonstrates that the videos selected are close to a uniform distribution; we verify this result for the other video resolutions but omit those results from the graph for clarity.

**Collection methodology.** We assess the QoE of each video stream by gathering various well-defined QoE KPIs [27] for every YouTube session. Each experiment starts by randomly selecting a video and streaming it twice, once with QUIC enabled (we confirm that QUIC is used via HTTP logs) and once with it disabled (TCP enabled). The order of the QUIC/TCP protocols is reversed for the next randomly chosen video to minimize bias introduced by CDN caching [2]. The resolution is set to automatic, and a Chrome extension is utilized to capture player events at 50 ms intervals. The player events are then transmitted to a local server for storage, while a Python script driving Selenium is used to capture HTTP logs. This process is repeated until every video in the dataset is streamed five times with each transport protocol. The collected QoE KPIs are as follows:

- **Average session resolution**: given $n$ available resolutions $R = \{R_1, R_2, \ldots R_n\}$ in a session, the fraction of time each resolution is viewed within the session is $P = \{P_1, P_2 \ldots P_n\}$; the average session resolution is given by $\sum_{i=1}^{n} P_i R_i$.
- **Initial buffering time**: the time from the instant when the video player connects to a server to the time when the first video frame is rendered and played. We filtered out all sessions with pre-roll advertisements.
- **Resolution changes**: the number of resolution switches per video. Frequent switches usually lead to unsatisfactory user experience [27].
- **Rebuffering events**: the number of times the playback pauses due to insufficient buffered video. Few or no rebuffering events are desired for better QoE.

**Video/audio chunks.** Video and audio chunks are the fundamental operational unit of ABR algorithms; therefore, our analysis focuses on network performance at the chunk granularity. To obtain the performance of these chunks, we first filter out all HTTP Network.requestWillBeSent events reported by Chrome. Then we ensure these requests contain ?videoplayback in their URL as well as video or audio as their MIME type. We keep track of the request_id of these events and obtain all Network.dataReceived events of the corresponding request_id. Finally, we group Network.dataReceived events by request_id to compute the size and performance metrics of each video/audio chunk.

Although such methods can only be applied if we have direct control over the client Chrome browser, previous work [8] has proposed methods to heuristically infer video/audio chunks based on the amount of data received between two HTTP requests. However, based on our HTTP logs, we found that the heuristic approach may no longer be viable because 45% of video and audio chunks are smaller than the previously defined threshold of 80 KB; these chunks could be as small as 4 KB. Additionally, contrary to previous literature [8], more than one chunk can be downloaded at a given time in the GEO satellite network. This requires us to model the chunk-downloading mechanism slightly differently.

**Chunk-level metrics.** In a GEO network, we have noted that video/audio chunks are primarily requested sequentially, but there are instances where a new request for additional chunks may be initiated before the previously requested chunks are fully received. To analyze the streaming behavior of the chunks, we utilize the following chunk-level metrics:

- **Chunk time to first byte (TTFB)**: the interval between the time a request is initiated and the first byte of data for the chunk is received. [23] identifies this metric to have an impact on chunk throughput; however, this should be differentiated from the idle time since it does not consider previous chunks.
- **Idle time**: the interval between the end of the previous chunk's transmission and the beginning of the following chunk's transmission. Due to a large accumulated playback buffer, YouTube may occasionally decide to pause chunk requests for an extended period of time (tens of seconds). This causes the chunk throughput to be low irrespective of network conditions. Therefore we only consider chunks that are requested within one second of the completion of the previous chunk. We also group chunks with negative idle time into larger chunks during our throughput analysis. This approach allows us to account for the overlapping download periods of these chunks.
- **Chunk download time**: the interval between when the first and final bytes of a chunk is received.
- **Chunk size**: the amount of data received in the requested chunk.
- **Throughput with idle time** ($T_{idle}$): the chunk throughput when idle time is taken into account. This metric demonstrates how imperfect request pipelining affects the throughput of chunks. Under shaped bandwidth, where buffer health remains relatively low, video chunk requests are made continuously and in a sequential manner. When these requests are perfectly pipelined, there should not be any idle time. The metric is given by:
$T_{idle} = \frac{Chunk\_Size}{Idle\_Time + Chunk\_Download\_Time}$
- **Throughput without idle time** ($T_{network}$): the chunk throughput when idle time is not taken into consideration. This demonstrates how effectively the chunk download is using the underlying network resource and is given by:
$T_{network} = \frac{Chunk\_Size}{Chunk\_Download\_Time}$

## 4 VIDEO STREAM PERFORMANCE

A key goal of our study is to understand the YouTube QoE received by users when they enable video bandwidth shaping rate to reduce data usage. As part of this analysis, we quantify any performance differences based on the use of gQUIC or TCP + PEP Proxy at a bandwidth shaping rate and a burst rate of 0.9 Mbps and 0.99 Mbps, respectively, which is the bandwidth shaping option supported by the production network.

We begin by analyzing the QoE KPIs. In Figure 2, we observe that the maximum average bit rate of 360p videos is only 0.63 Mbps, which is approximately 70% of the traffic shaping rate. As a result, we expect the network to support videos at 360p or lower quality seamlessly, i.e., without rebuffering events. However, contrary to our expectations, we find that only 72% of TCP and 82% of gQUIC sessions experience zero rebuffering events (we include this graph in the

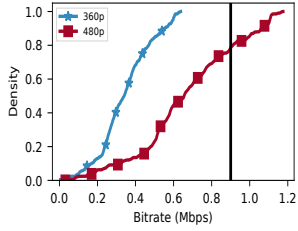
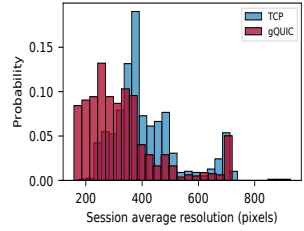
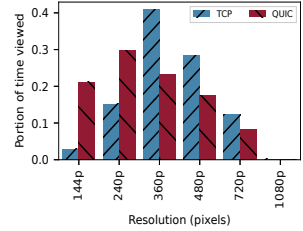
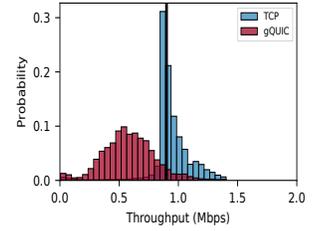

Figure 2: Average video bit rates for each resolution. Vertical line is 900 Kbps.

Figure 3: Session average vertical pixel height.

Figure 4: Percent of time viewed at each resolution.

Figure 5: Throughput without considering idle time. Vertical line is 900 Kbps.

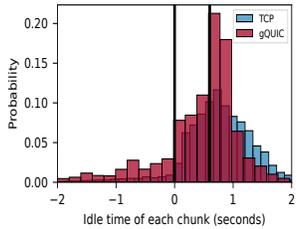
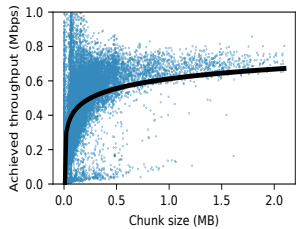

Figure 6: Idle time. Vertical lines are 0 ms and 600 ms idle time.

Figure 7: Chunk size vs throughput for QUIC.

Appendix for completeness). Notably, the resolution achieved by gQUIC sessions is significantly lower than that of TCP sessions. As shown in Figure 3, the median average session resolution for TCP is 380p, whereas for gQUIC the median average session resolution is only 299p. Specifically, 18% of the TCP streaming time is at a resolution less than 360p, while 50% of the gQUIC streaming time is less than 360p, as shown in Figure 4. Surprisingly, these metrics indicate that neither the gQUIC nor TCP sessions consistently meet our QoE expectations, with gQUIC, in particular, struggling to stream at 360p.

We next examine the initial buffering delay and number of resolution changes. The number of resolution changes for TCP and gQUIC are similar, with mean values of 3.57 and 3.77 per session, respectively. The mean initial buffering delay of TCP is 8.77 seconds while gQUIC is 10.74 seconds. These graphs are included in the Appendix.

In summary, the results show that gQUIC has fewer re-buffering events, at the cost of lower average resolution, more resolution changes, and greater initial buffering delay. Further, neither gQUIC nor TCP is able to stream seamlessly at 360p when bandwidth shaping is utilized. We hypothesize that there are suboptimal transport and application layer operations that manifest in unexpectedly poor performance in long-latency networks, preventing both gQUIC and TCP from fully utilizing available network resources and achieving better performance. In the following section, we investigate these inefficiencies more deeply.

## 5 DIAGNOSIS OF SUB-OPTIMAL QOE

Section 4 showed that both TCP and gQUIC fail to deliver expected QoE KPIs, with gQUIC, in particular, struggling to provide higher resolution. In this section, we dig deeper into these results to understand this performance anomaly.

### 5.1 Insufficient network resource utilization

Figures 2 and 4 demonstrated that sufficient bandwidth was available to stream at 360p, yet 50% of the time, gQUIC streamed at a lower resolution. Our goal is to understand why gQUIC is unable to utilize available network resources efficiently. We begin with the throughput of individual chunks when the idle time is not considered, depicted in Figure 5. The figure shows that TCP clearly forms a peak around 900 Kbps, the shaped bandwidth, whereas gQUIC's throughput widely varies well below 900 Kbps. The mean and median of TCP $T_{network}$ are 0.93 Mbps and 0.91 Mbps, respectively, while the mean and median of gQUIC $T_{network}$ are 0.58 and 0.57 Mbps.

Both TCP and gQUIC use BBR as their congestion control algorithm. BBR continuously probes bottleneck bandwidth and propagation delay to determine the sending rate; however, we have seen that gQUIC does not send at a rate that saturates the capacity of the link. Therefore, we hypothesize that the BBR algorithm performs differently with gQUIC and TCP in GEO networks, likely as a result of PEP in the network architecture. Specifically, gQUIC BBR experiences the full 600 ms of latency across the satellite link, whereas TCP BBR experiences only the low-latency link (typically less than 100 ms round trip time) between the YouTube server and PEP server proxy. The difference in this propagation delay may cause the YouTube server to incorrectly estimate the available bandwidth and send data at a rate that never fully saturates the link when gQUIC is used. Prior work, such as [31] and [4], has also observed that the combination of TCP+PEP+BBR achieves higher throughput than gQUIC+BBR. Relatedly, [31] showed that the slow-start phase of gQUIC+BBR is the cause of the lower throughput; gQUIC+BBR requires up to 10 seconds to reach maximum throughput, whereas TCP takes less than one second. In our analysis, we observe that $T_{network}$ of gQUIC is still variable well below 900 Kbps even for video chunks requested 30 seconds or more after the session start. As a result, the start-up phase is not the only factor that affects the performance of gQUIC+BBR in our test network.

## 5.2 YouTube scheduling inefficiencies

Prior work has shown that many YouTube videos are not watched to completion [22]. As a result, YouTube tries to reduce unnecessary bandwidth consumption by limiting the number of chunks downloaded in advance. Our analysis has shown that it is possible for the currently requested chunk to start transmission before the completion of the previous chunk. Ideally, the YouTube ABR should learn the RTT of the network and request the next chunk slightly before the completion of the current chunk. This would minimize the amount of time spent idling by the network and therefore increase network utilization. However, as shown in Figure 6, we observe that more than 80% of the chunks experience idle time with both TCP and gQUIC. Interestingly, we can also observe that a peak is formed around 600 ms. This suggests that the high propagation delay of the GEO network is not considered when making chunk requests.

In order to assess the impact of this idle time on smaller chunks whose throughput is dominated by idle time, we analyze the relationship between the size of the chunk downloaded and the throughput achieved for both TCP and gQUIC. Figure 7 shows a strong positive Pearson correlation of 0.49 of the two variables for gQUIC; the black fitted line can visually confirm this. We verify these two variables are also highly correlated for TCP with a Pearson statistic of 0.36. Similar trends are observed in [15] for 4G and WiFi networks. This leads us to infer that these pipelining inefficiencies significantly reduce the throughput of smaller chunks. In addition, shaped traffic on GEO networks is disproportionately impacted due to their lower bandwidth and higher latency. According to [20], YouTube tends to request smaller chunks when bandwidth is lower. We observe that more than 50% of all chunks in our sessions are less than 0.1 MB; this, in turn, further exacerbates the problem. It is worth noting that the chunk-requesting inefficiency observed in our study may be specific to low-bandwidth high-latency scenarios.

To investigate this hypothesis, we compare the impact of idle time in the GEO network with that on our campus network, which is both high bandwidth and low latency. We stream the same 2,080 videos to a desktop in our campus research lab and find that the median idle time for both TCP and gQUIC is short, approximately 15 ms (supporting the assumption of almost ideal sequential chunk requesting). Moreover, given the abundance of campus network capacity, the idle time has no practical impact on QoE in this setting, with $T_{idle}$ exceeding 60 Mbps. In contrast, in the shaped GEO network, the median $T_{idle}$ for TCP and gQUIC is approximately 0.54 Mbps (0.6 $T_{network}$) and 0.4 Mbps (0.7 $T_{network}$), respectively. In summary, TCP experiences a 40% throughput reduction due to idle time, even with the link being saturated during transmission, while gQUIC suffers a 30% reduction in throughput and is unable to fully utilize the link capacity during transmission.

## 6 RELATED WORK

Prior literature has studied the QoE of ABR streaming and YouTube [9, 19, 21, 32]. However, far fewer studies have investigated video streaming QoE in operational GEO networks. In [7], a framework that facilitates live 4k video streaming over a 5G core network via a GEO satellite backhaul is proposed. However, the focus of this work is live video streaming, as opposed to non-real-time video streaming in our study. [30] proposes a caching framework that improves video streaming QoE within GEO satellite networks. The limitations of this study include the utilization of an academic DASH player and the investigation of only a single video. Distorted versions of videos are generated in [33] by adjusting QoS parameters such as packet loss. The distorted videos are replayed to volunteers, and the corresponding Mean Opinion Score is recorded. This work uses an emulated satellite link, and the impact of ABR is not considered. Extensive prior work has analyzed transport protocol performance; however, these studies utilize a low-latency link. For instance, [28] uses a network emulator to shape home and mobile networks to 1 Mbps to study YouTube streaming QoE. In this environment, the authors conclude there is no meaningful difference between TCP and QUIC performance. Through the use of an academic DASH player in a terrestrial network and only one video, [3] concludes that the QUIC protocol does not improve QoE. Finally, the page load time difference between TCP and QUIC is studied in [18].

## 7 CONCLUSION

Our study characterizes the QoE KPIs of YouTube in a production GEO satellite network with 900 Kbps traffic shaping. We find that, despite the bit rate of the 360p videos being less than the shaped bandwidth, the performance of these video streams is sub-optimal. Our work highlights the challenge of employing traffic shaping as an effective means to control video resolution/data-usage on high latency networks, and the importance of accounting for network delay in chunk size and request timing for high quality video streaming. Many of the populations that stand to gain Internet access over the coming years will do so through non-traditional network architectures, including GEO networks. Application designers and content providers must consider a wide variety of network types and characteristics in their product, protocol design and content provisioning strategies to avoid unanticipated performance anomalies.

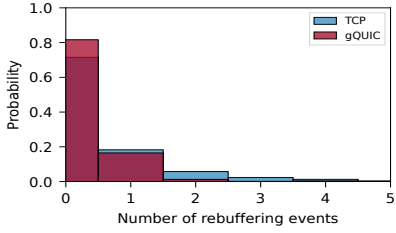

Figure 8: Number of rebuffering events.

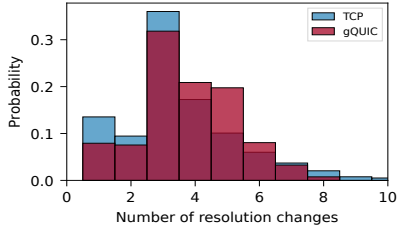

Figure 9: Number of resolution changes.

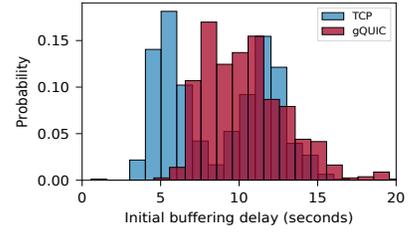

Figure 10: Initial buffering time.

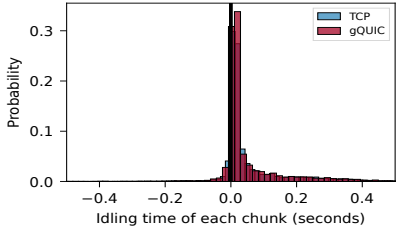

Figure 11: Idle time of campus network.

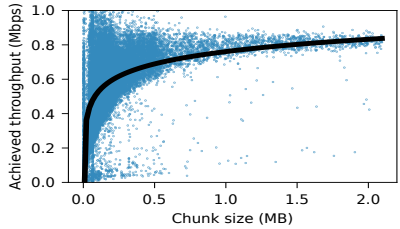

Figure 12: Chunk size vs throughput over TCP.

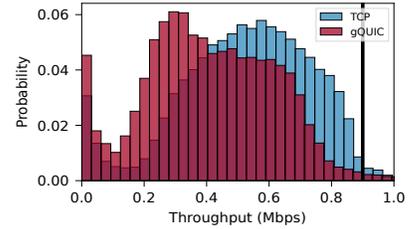

Figure 13: Achieved throughput. Vertical line is 900 Kbps

## A APPENDIX

### A.1 Ethical Considerations

Although our work involves HTTP log analysis on an operational GEO satellite network, our work is not human subjects research. At no point is any data collected from the customers of the network. We collect and analyze only our own experimentally generated traffic.

### A.2 Supplementary Results

In this section we include some additional, supplementary graphs that were briefly described in the main body of the paper. Figure 8 shows the number of rebuffering events for TCP and QUIC. We observe that 28% of TCP and 18% of QUIC sessions have at least one rebuffering events. Figure 9 shows that the number of resolution changes for TCP and QUIC are similar, with mean values of 3.57 and 3.77, respectively. The initial buffering delay is shown in Figure 10; the mean delay of TCP is 8.77 seconds and of QUIC is 10.74 seconds. These QoE KPIs, combined with the average resolution metric mentioned in Section 4, indicate that both TCP and QUIC video sessions suffer from rebuffering events and/or low resolution. Consequently, we conclude that the QoE falls below our expectations based on average video bit rates. The median idle time for both TCP and gQUIC was short in our campus network experiment, around 15 ms, as shown in Figure 11. This suggests a correlation between idle time and the round trip propagation delay of the network, providing further evidence that pipelining is sub-optimal in a GEO satellite network. The correlation between achieved throughput ($T_idle$) and chunk size for TCP, similar to QUIC, is illustrated in Figure 12. The Pearson statistic for this correlation is 0.36. Finally, in Figure 13 we see that the $T_idle$ of TCP outperforms that of QUIC in GEO networks with a median throughput of 0.54 Mbps, compared to QUIC's median of 0.4 Mbps. Critically, however, neither reach the shaped bandwidth rate.